\documentclass[conference,a4paper]{IEEEtran}

\usepackage{cite}
\usepackage{array}
\usepackage{xcolor}
\usepackage{mdwtab}
\usepackage{amssymb}
\usepackage{makeidx}
\usepackage{hyperref}
\usepackage{graphicx}
\usepackage{textcomp}
\usepackage{placeins}
\usepackage{epstopdf}
\usepackage{enumitem}
\usepackage{tabularx} 
\usepackage{algorithmic}
\usepackage{subfigure}
\usepackage{booktabs,caption}
\usepackage{psfrag, graphicx}
\usepackage[latin1]{inputenc}
\usepackage{amsmath,amssymb,amsfonts}
\usepackage[flushleft]{threeparttable}
\usepackage{amsfonts,amsmath,amssymb,bbm}
\usepackage{graphicx,color,epsfig,rotating}

\newcolumntype{C}[1]{>{\centering\arraybackslash}m{#1}}

\usepackage{nopageno}
\begin{document}



\title{SPARC-LoRa: A Scalable, Power-efficient, Affordable, Reliable, and Cloud Service-enabled LoRa Networking System for Agriculture Applications}
 
\author{Xi Wang$^{1}$, Bryan Hatasaka$^{1}$, Zhengyan Liu$^{1}$, Sayali Tope$^{1}$, Mohit Karkhanis$^{1}$, Seungbeom Noh$^{1}$, Farhan Sium$^{1}$, Ravi V. Mural$^{3}$, Ling Zang$^{2}$, James Schnable$^{3}$, Carlos Mastrangelo$^{1}$, Hanseup Kim$^{1}$,  and Mingyue Ji$^{1}{*}$

\thanks{The authors are with the Department of Electrical and Computer Engineering, University of Utah, Salt Lake City, UT 84112, USA. (e-mail: leo.wang@utah.edu, and mingyue.ji@utah.edu)} }

\author{
    \IEEEauthorblockN{ Xi Wang$^{1}$, Bryan Hatasaka$^{1}$, Zhengyan Liu$^{1}$, Sayali Tope$^{1}$,
    Mohit Karkhanis$^{1}$, Seungbeom Noh$^{1}$, \\
    Farhan Sium$^{1}$, Ravi V. Mural$^{3}$, Hanseup Kim$^{1}$, Carlos Mastrangelo$^{1}$, Ling Zang$^{2}$, \\ James Schnable$^{3}$, and Mingyue Ji$^{1}$ }
    
	\IEEEauthorblockA{$^1$Department of Electrical and Computer Engineering, University of Utah, \\
                    $^2$Department of Material Science and Engineering, University of Utah, \\
	              $^3$Department of Agronomy and Horticulture, University of Nebraska-Lincoln  \\ }
		               
    Email:\{$^1$\{leo.wang, u1028471, u1225560, sayali.tope, mohit.karkhanis, moses.noh, u1369555,  \\
            hanseup.kim, carlos.mastrangelo, mingyue.ji\}@utah.edu,
            $^2$lzang@eng.utah.edu, $^3$\{rmural2, schnable\}@unl.edu \}
            
}

\maketitle

\thispagestyle{plain}
\pagestyle{plain}

\vspace{-0.5cm}

\begin{abstract}
With the rapid development of cloud and edge computing, 
Internet of Things (IoT) applications have been deployed in various aspects of human life. 
In this paper, we design and implement a holistic LoRa-based IoT system with LoRa communication capabilities,  
named SPARC-LoRa, which consists of field sensor nodes and a gateway connected to the Internet. 
SPARC-LoRa has the following important features. First, the proposed wireless network of SPARC-LoRa is even-driven and using off-the-shelf microcontroller and LoRa communication modules with a customized PCB design to integrate all the hardware. This enables SPARC-LoRa to achieve low power consumption, long range communication, and low cost. With a new connection-based upper layer protocol design, the scalability and communication reliability of SPARC-loRa can be achieved. Second, an open source software including sensor nodes and servers is designed based on Docker container with cloud storage, computing, and LTE functionalities. In order to achieve reliable wireless communication under extreme conditions, a relay module is designed and applied to SPARC-LoRa to forward the data from sensor nodes to the gateway node. The server side software platform consists of 1) MQTT protocol publishing measured data from the gateway node to the cloud servers using LTE connections; 2) InfluxDB database for data storage; and 3) Grafana data web server for accessing the measured data from any platforms or web browsers. The system design and implementation is completely open source and hosted on the DigitalOcean Droplet Cloud. Hence, the proposed system enables further research and applications in both academia and industry. The proposed system has been tested in real fields under different and extreme environmental conditions in Salt Lake City, Utah and the University of Nebraska-Lincoln. The experimental results validate the features of SPARC-LoRa including low power, reliability, and cloud services provided by SPARC-LoRa.
\end{abstract}


\section{Introduction} \label{section: introduction}
The Internet of Things (IoT) is a rapidly growing field that has distinct advantages when integrated into current traditional production processes and industries. One such industry that can benefit from IoT is agriculture \cite{yang2020design,chanwattanapong2021lora,gak2022lora}. Traditionally, farmers rely on physical inspection of crops to identify problems with weed or insect outbreaks. These inspections can be expensive, costly, and slow. These problems can be identified quickly and crops can be saved by designing and implementing a low power, long range chemical gas detection wireless sensor network. The authors in \cite{bandyopadhyay2020iot} describe that the technology and engineering has gone into developing the IoT aspect of such wireless sensor networks. Low power and event driven techniques are utilized by a cooperating fabricated sensor and microcontroller circuits to consume power only when required. 
In \cite{zhou2019design}, the authors apply \underline{lo}ng \underline{ra}nge (LoRa) communication technique to communicate information to a gateway device over wide distances and variable environments of crop fields. 
Traditional mesh networks are not practical for this application as higher power consumption results in a need to replace sensor batteries before the end of a crop season \cite{ali2019zigbee}. 


\begin{figure}
    \centering
    \includegraphics[width=75mm]{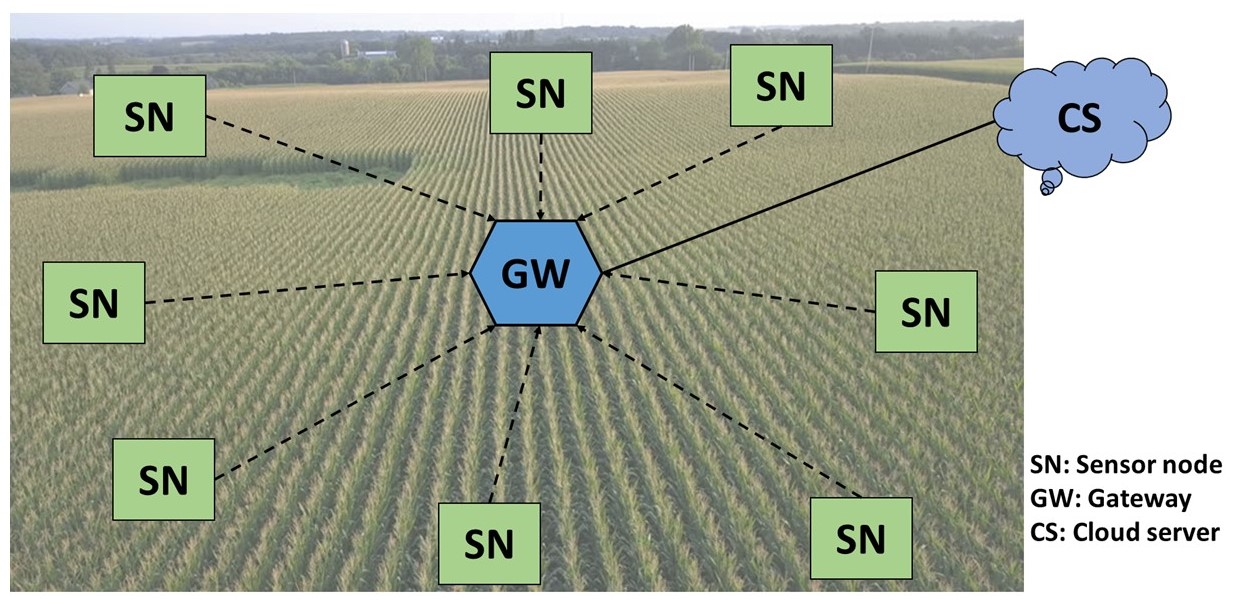}
    \caption{Overview of SPARC-LoRa Networking System.}
    \vspace{-0.3cm}
    \label{fig: system architecture}
\end{figure}

The agriculture IoT systems have the following stringent requirements. First, the sensor nodes deployed in the field have to be low power in order to guarantee they can last for at least a season without changing batteries. Second, long range communication capability is required due to the large size of a normal field. Third, the systems need to be reliable, meaning that all the data generated by the sensor nodes can be transmitted successfully and viewed by any platforms. Fourth, the size of the sensor nodes needs to be small since it may be ideal to mount it on the leaves of crops. However, the systems do not need to have low communication latency since the data generated in the field is not urgent and should be received within approximately a day. Moreover, the systems do not require high transmission rate and high computation capability at the sensor nodes. Based on these requirements, we propose a \underline{s}calable, \underline{p}ower-efficient, \underline{a}ffordable, \underline{r}eliable, and \underline{c}loud service-enabled LoRa network networking system (SPARC-LoRa) for agriculture applications in this paper. In particular, this system is designed specifically for transmitting the detected gas (e.g, hexanal, indole) information of plants to determine abnormal crops production at an early stage.  SPARC-LoRa can satisfy all the requirements listed above. In addition, possible cloud computing based service (e.g., data analysis) is available for future development, since the server side software of SPARC-LoRa is hosted on a cloud service platform. The overview of this system is shown in Fig.\ref{fig: system architecture}. Once sensor nodes detect some events (e.g., the density of hexanal is above a declared threshold), the transmission from the sensor nodes to the gateway will be triggered and then the gateway will publish the data to the Internet (e.g., some cloud servers). 

The contributions of this paper are as follows. 
\begin{enumerate}
    \item We design and implement SPARC-LoRa for agriculture applications. SPARC-LoRa is scalable, power-efficient, affordable, and reliable. And it enables cloud services. All the components of SPARC-LoRa are low cost. 
    \item On the sensor nodes side, the low power requirement is achieved by our customized PCB that integrates a power efficient microcontroller and off-the-shelf LoRa communication module. The scalability and reliability requirements are achieved by our new connection-based upper layer networking protocol, the relay module, and the long range communication capability of LoRa module. 
    \item on the sever side, the designed software architecture consists of 1) MQTT protocol publishing measured data from the gateway nodes to the cloud servers using LTE connections; 2) InfluxDB database for data storage; and 3) Grafana data web server for accessing the measured data from any platforms or web browsers. The system design and implementation is completely open source and hosted on the DigitalOcean Droplet Cloud. 
    \item SPARC-LoRa has been tested in real fields under different and extreme environmental conditions in Salt Lake City and the University of Nebraska-Lincoln. The testing results validate the features of SPARC-LoRa. 
\end{enumerate}


\section{System Components}
SPARC-LoRa consists of the hardware components and the software components, which will be discussed in the following. 

\subsection{Hardware Components}\label{hardware components}
Hardware is the platform for running software programs and performing data communications. 
In SPARC-LoRa, hardware components mainly include microcontroller unit (MCU), LoRa communication modules, and LTE communication modules to achieve the wireless communication and publish data online.

\subsubsection{MCU}
\begin{figure}
    \centering
    \includegraphics[width=65mm]{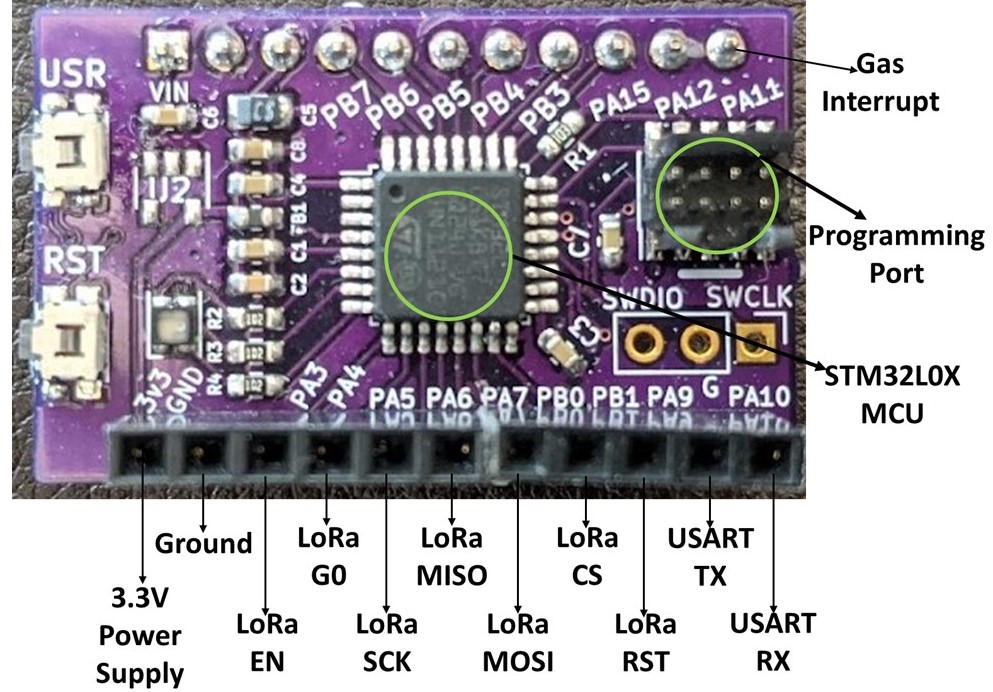}
    \caption{Customized Microcontroller Unit.}
\vspace{-0.3cm}
    \label{fig: MCU_pinout}
\end{figure}

The ARM based Microcontroller STM32L051K8 is chosen for SPARC-LoRa for its low power consumption. This MCU has sleep mode and active mode. The MCU is working under the active mode during the data transmission, and it will turn into sleep mode when the transmission is finished. The MCU consumes power in {millliwatts ($m W$) or microwatts ($\mu W$)}. 
The MCU is integrated into a customized PCB and the corresponding functions of the pins are shown in Fig.~\ref{fig: MCU_pinout} and Table~\ref{table: LoRa unit pin out function description}. The design of the PCB will be discussed in more details in Section~\ref{design novelties}. 

\begin{table}[h]
    \centering
    \caption{MCU Pin Out Function Description}
    
    \begin{tabular}{|C{1.2cm} | C{1.5cm} | C{4.8cm}|}
    \hline
    \textbf{Pin Label}      &       \textbf{Pin Out}        &       \textbf{Function Description }        \\
    \hline
    3v3                     &       3.3V Power Supply       &       Power the LoRa module       \\
    \hline
    GND                     &       Ground                  &       Ground for power and logic of LoRa module       \\
    \hline
    PA3                     &       LoRa EN                 &       Enable pin of the regulator     \\
    \hline
    PA4                     &       LoRa G0                 &       GPIO 0 pin or IRQ pin for interrupt request notification        \\
    \hline
    PA5                     &       LoRa SCK                &       SPI clock pin       \\
    \hline
    PA6                     &       LoRa MISO               &       Microcontroller in serial out pin for data sent from the radio to the processor        \\
    \hline
    PA7                     &       LoRa MOSI               &       Microcontroller out serial in pin for data sent from the processor to the radio        \\
    \hline
    PB0                     &       LoRa CS                 &       Chip select pin     \\
    \hline
    PB1                     &       LoRa RST                &       Reset pin for the radio     \\
    \hline
    PA9                     &       USART TX                &       Configuration pin to set ID/type of the unit, connected to RX pin of configuration device       \\
    \hline
    PA10                    &       USART RX                &       Configuration pin to set ID/type of the unit, connected to TX pin of configuration device       \\
    \hline
    PA11                    &       Gas Interrupt           &       Gas interrupt input to LoRa unit        \\
    \hline
    \end{tabular}

    \label{table: LoRa unit pin out function description}
    
\end{table}


\subsubsection{LoRa Module}
The LoRa module is long range radio communication device working in relatively low power. 
The communication between the \underline{s}ensor \underline{n}odes (SNs) and the \underline{g}ate\underline{w}ay (GW) is implemented using LoRa communication. The selected LoRa development module is Adafruit RFM95W, which uses Semtech SX1276 radio internally. This module has developed a library for ARM based MCU and works on the license-free ISM (915 MHz) frequency band for data transmission. In addition, a simple wire antenna is used for the LoRa module. The LoRa module with the wire antenna is shown in Fig.~\ref{fig: Lora module}. The integrated MCU and LoRa module via the single customized PCB is called the LoRa unit in the rest of this paper. 



\subsubsection{LTE Module}
The communication between the GW and the Internet (cloud server) is implemented using LTE Cat M1,\cite{hsieh2018experimental} which is a cellular LTE technology specifically designed for IoT applications. 
The reasons that LTE Cat M1 is feasible for SPARC-LoRa include two folds. First, LTE Cat M1 is available in the targeted fields in both Salt Lake City and the University of Nebraska-Lincoln. Second, although the power consumption of the LTE model is much higher than the LoRa module, a much bigger battery may be installed since the size of the GW could be large. In addition, it is worth mentioning that the signal strength of LTE varies in different locations, it needs to find a location with relatively strong signal to deploy the GW. 
The specific LTE module used in the proposed system is Boron LTE from Particle Industries Inc. This LTE module is shown in Fig.~\ref{fig: Gateway components}.



\subsection{Software Components}\label{software components}

Two software components are implemented in SPARC-LoRa as follows. 

\subsubsection{SN Side Software}

SN side software is written in the MCU via the configurator (see Fig.~\ref{fig: Sensor node components}) programs the LoRa unit. Arduino IDE is used for programming and provides the interface between software and hardware. The programming is implemented using Arduino programming language in Arduino IDE while the library is written in C++ programming language. The software implements four functions 
whenever it is applicable including MCU sleeping mode, data transmitting mode, data receiving mode, and data forwarding model in the relay case. 


\subsubsection{Server Side Software}\label{server side software}
Server side software will be discussed as follows.  

\begin{itemize}
    \item MQTT is a lightweight, publish-subscribe, and widely used protocol for IoT messaging. To achieve LTE communication to Internet, we use MQTT protocol to send data from GW to Internet in SPARC-loRa. MQTT server publishes data to any subscribers including network clients, phones, and intermediate program. 
    \item An intermediate program is designed and implemented in Python programming language to store the data coming from MQTT server in InfluxDB. 
    \item We use InfluxDB for the database in SPARC-LoRa. InfluxDB is a time series database which stores the data based on storage time, and it is optimized for logging.
    \item Garafana is a web-based data viewer. In the proposed system, Grafana is connected to InfluxDB source and allows collected data to be read from anywhere and viewed from any platforms including web browsers. 
\end{itemize}

The above software is hosted on a DigitalOcean Droplet Cloud and runs in a Docker container, which is convenient for portability, maintenance, and versioning. In addition, it is possible to add cloud computing related services if needed.

\section{System Design}

In this section, we will discuss the details in the design of SPARC-LoRa. The system architecture of SPARC-LoRa is shown in Fig.~\ref{fig: System model}. 

\subsection{System Modules}\label{system modules}
The proposed network has three physical modules including SN module, GW module, and relay node (RN) module. Each module is a transceiver which includes an MCU and a LoRa module. The 
GW module has an Internet module to support the data publication online, store data to online database, and provide the interface for users to view the data.

\subsubsection{SN Module}
Each SN is physically connected to the chemical sensor, which is used to detect the gases, to receive the gas interrupt (see Fig.~\ref{fig: MCU_pinout}). The MCU will be triggered when there is an gas interrupt, then the MCU initializes the data transmission.


\subsubsection{GW Module}
The GW has a MCU, a LoRa Unit, and LTE modules. Its functionalities include receiving data from either SNs and RNs (see next paragraph), returning an acknowledgement to SNs, and publishing the received data to the Internet. 


\subsubsection{RN Module}
\begin{figure}[htbp]
    \centering
    \includegraphics[width=80mm]{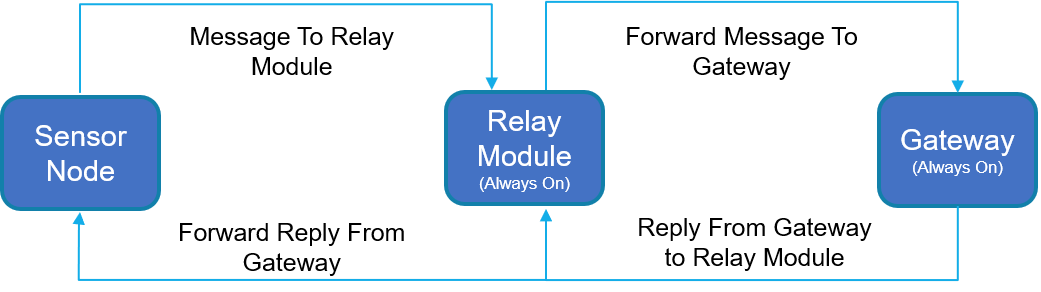}
    \caption{Relay Module Functionality.}
    \label{fig.Relay_function}
\end{figure}

The RN module is activated when the direct communication between the SN and GW is not available. In this case, the RN module forwards the data from SN to the GW, and vice versa. The RN Module is shown in Fig.~\ref{fig.Relay_function}.

\subsubsection{Internet Module}
\begin{figure}[htbp]
    \centering
    \includegraphics[width=85mm]{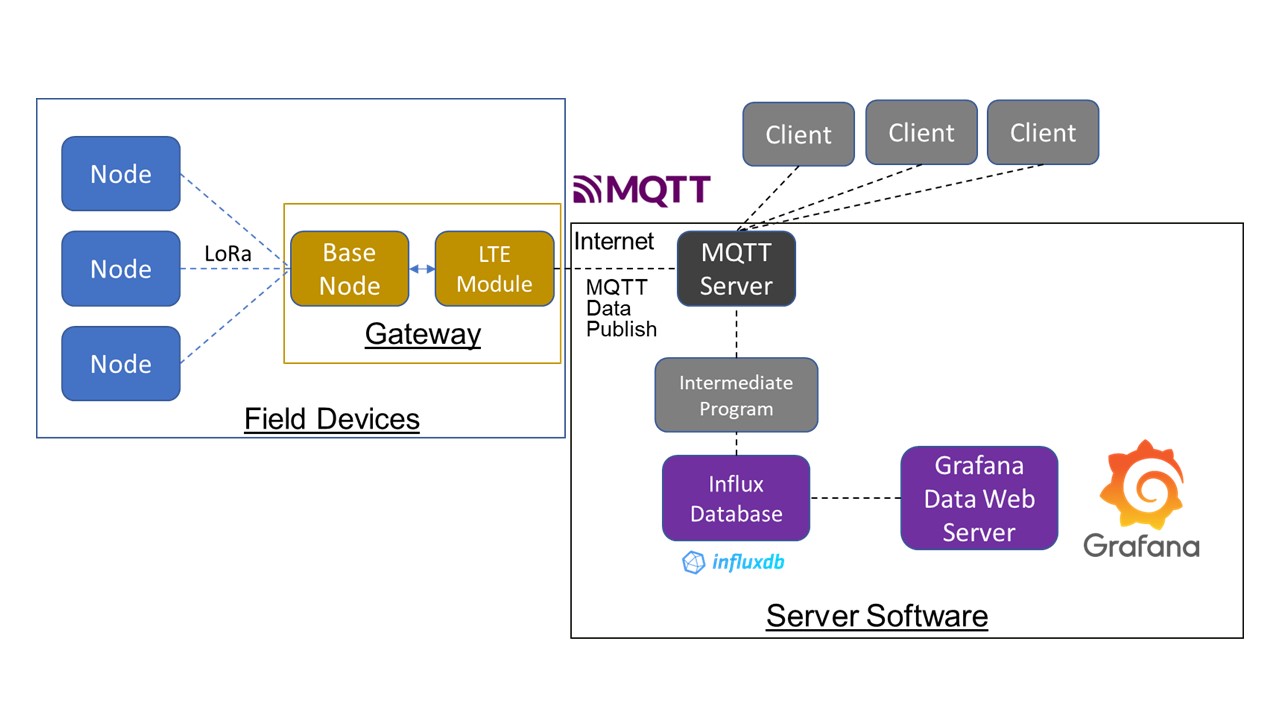}
    \caption{Overview of the Architecture of SPARC-LoRa.}
    \label{fig: System model}
\end{figure}

Internet module integrates the MQTT server, an intermediate program, the InfluxDB database, and the Grafana server in order to allow users to view the data sent from the SNs. 
As shown in the right part of Fig.~\ref{fig: System model}, Internet module transmits data from GW to the Internet and allows that data is viewable to any users who have access. MQTT protocol is used to publish client data from local to MQTT server on Internet. In SPARC-LoRa, each LTE module is a client of the MQTT server, therefore, the LTE data on GW is published to the MQTT server. An intermediate program written in Python is used to store MQTT server data in InfluxDB in between the MQTT server and InfluxDB, since he MQTT server is not able to store the data. The intermediate program is running in a virtual machine and always ready to store the MQTT server data. Data that stores in InfluxDB is finally viewable on Grafana data web server to any users who have access from any platforms or browsers.

\subsection{Field Devices}\label{Field Devices}
\begin{figure}
    \centering

    \subfigure[Sensor Node]{\includegraphics[width=0.48\linewidth]{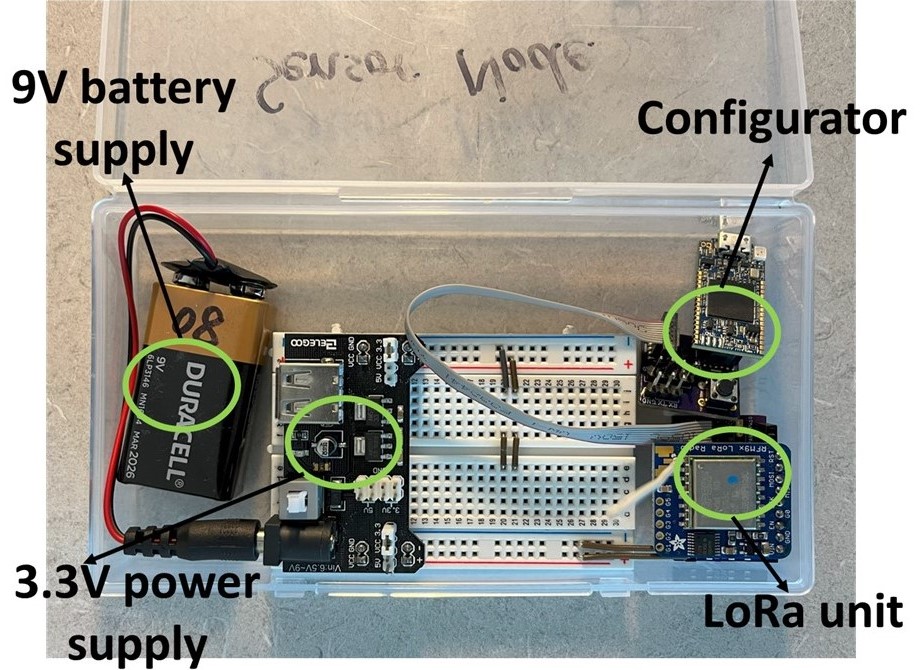}\label{fig: Sensor node components}}
    \hspace{0.2mm}
    \subfigure[Gateway]{\includegraphics[width=0.48\linewidth]{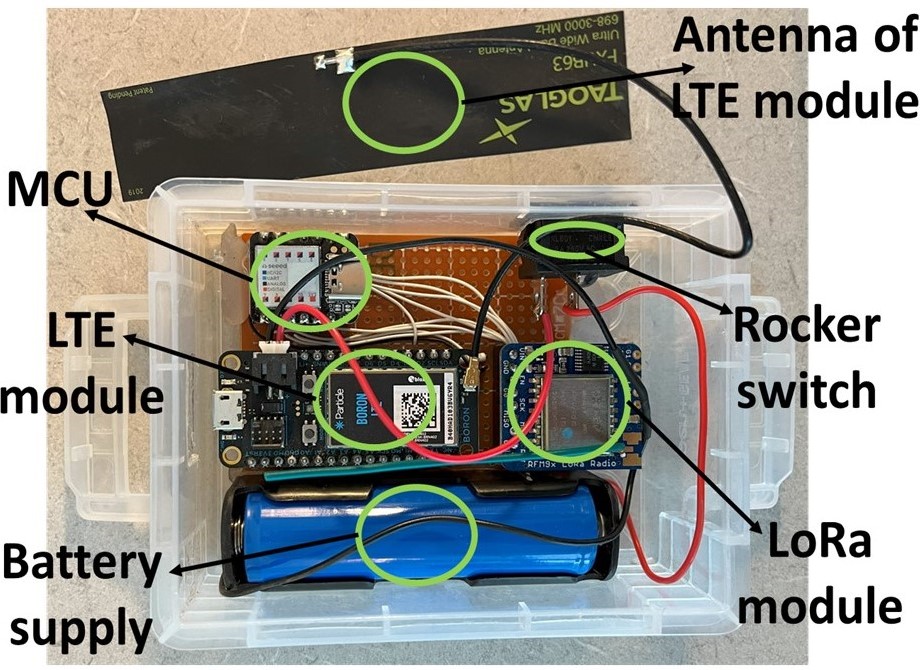}\label{fig: Gateway components}}
    \hspace{0.2mm}
    \subfigure[Relay Node]{\includegraphics[width=0.48\linewidth]{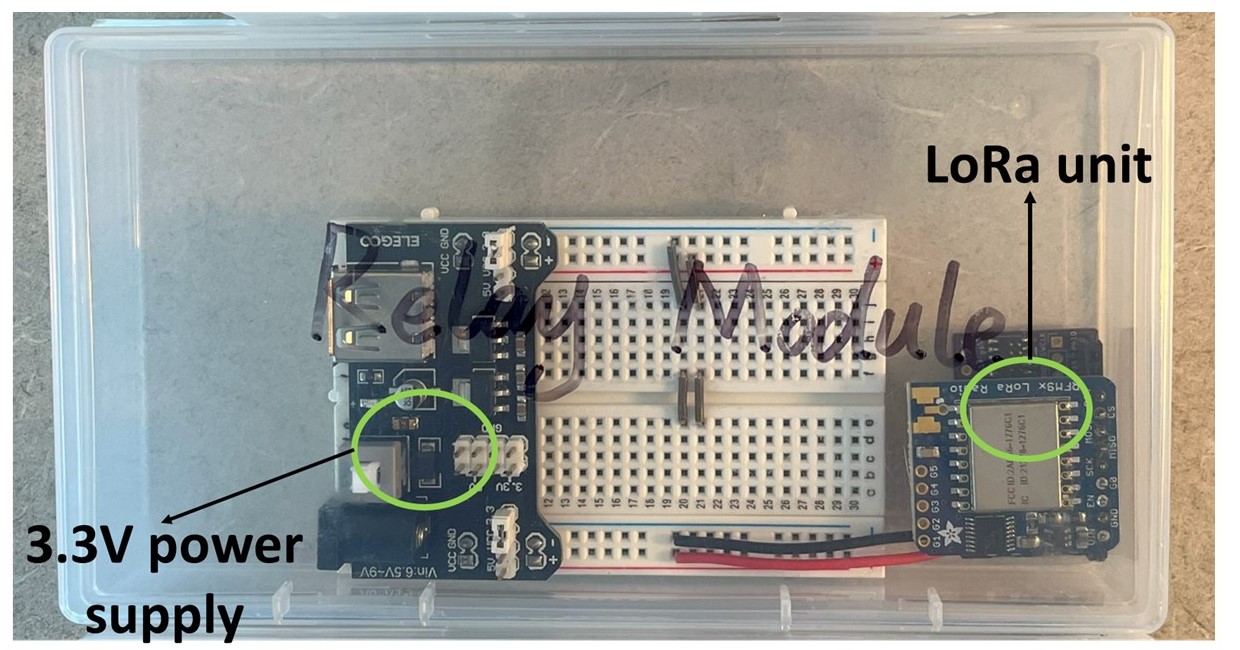}\label{fig: Relay noode components}}
    \hspace{0.2mm}
    \subfigure[Adafruit RFM95W LoRa Radio Transceiver]{\includegraphics[width=0.48\linewidth]{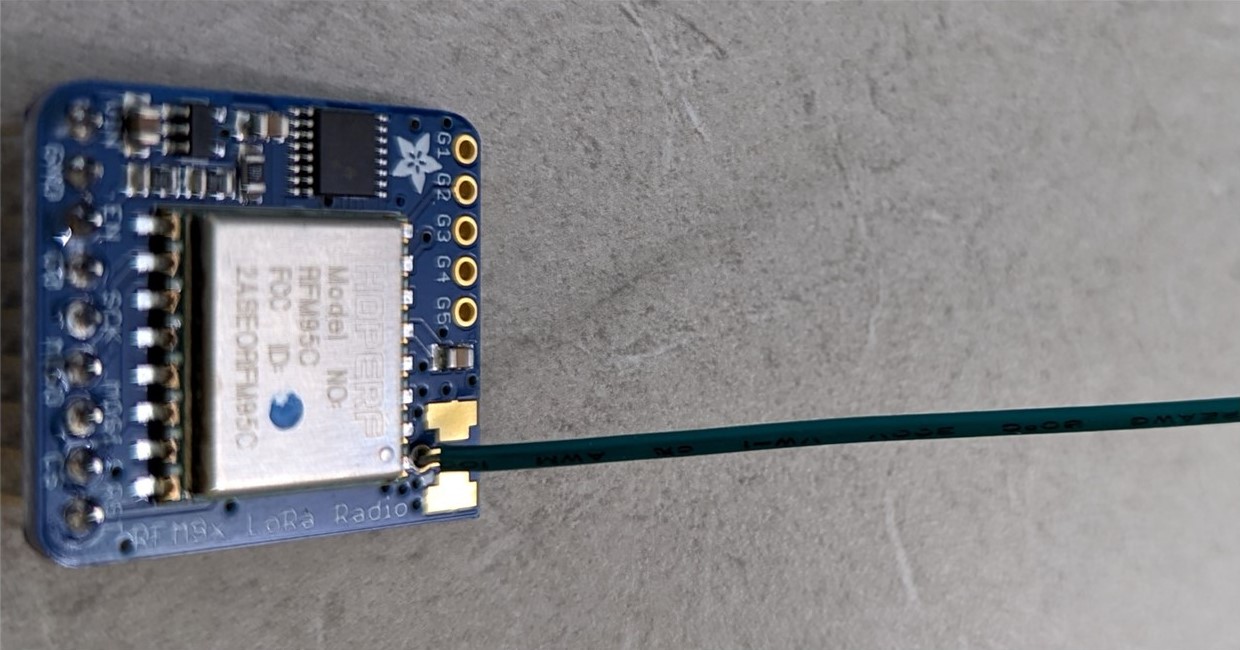}\label{fig: Lora module}}

  \caption{Wireless Sensor System Physical Components.}
  \vspace{-0.3cm}
  \label{fig: Wireless sensor physical system components}
\end{figure}

Field devices are SNs, GW, and RNs, which are physically deployed in the testing field. Each SN includes 
a battery supply, a power supply, a configurator, and a LoRa unit as shown in \ref{fig: Sensor node components}.  
The GW includes all the components of the SN, the LTE module, and the rocker switch which is for turning on/off the power. 
The RN has the exact same components of the SN but may be installed a bigger battery since there will be no chemical sensor in it. 

Each SNs 
is powered by a 3.7 Lithium poly battery, which is directly connected to the processing unit (part of the chemical sensor and not shown in Fig.~\ref{fig: Wireless sensor physical system components}), and the LoRa unit is powered by the processing unit through power and ground wires. There is 
another wire connected between the processing unit and LoRa unit for gas interrupt purpose. The LoRa unit is always on sleep mode if there is no data transmission to reduce power consumption, and is only woken up by the interrupt received from the processing unit to start the data transmission. Whenever the 
gas arrives to the declared threshold of the chemical sensor, the processing unit amplifies the incoming signal from the chemical sensor and send an interrupt to the LoRa unit. Then the MCU wakes up from sleep mode and 
wakes up the LoRa module, and the MCU initializes the LoRa transmission from the SN to the GW. 
The LoRa module communicates with the MCU over serial peripheral interface (SPI) protocol and the MCU communicates with the LTE module over universal asynchronous receiver-transmitter (UART) protocol. GW module forwards the data 
to server side software.


\subsection{Design Details and Novelties}\label{design novelties}

The design details and the novelties of SPARC-LoRa are as follows.

\subsubsection{Physical Layer}

RadioHead Packet Radio library for embedded microprocessors is used for the physical layer of LoRa module \cite{radiohead}. It allows changing the two important parameters including spreading factor and coding factor. These parameters can control the tradeoff between the communication distance and the transmission rate. In particular, the largest spreading factor ($12$) and the smallest coding rate ($4/8$) achieve the longest transmission distance and the smallest transmission rate. Due to the reliability requirement, we choose the spreading factor of $12$ and coding rate of $4/8$. In addition, the RadioHead also does a simple collision detection and backoff window-based Medium Access Control (MAC) layer protocol. 




\begin{figure}[htbp]
    \centering
    \includegraphics[width=60mm]{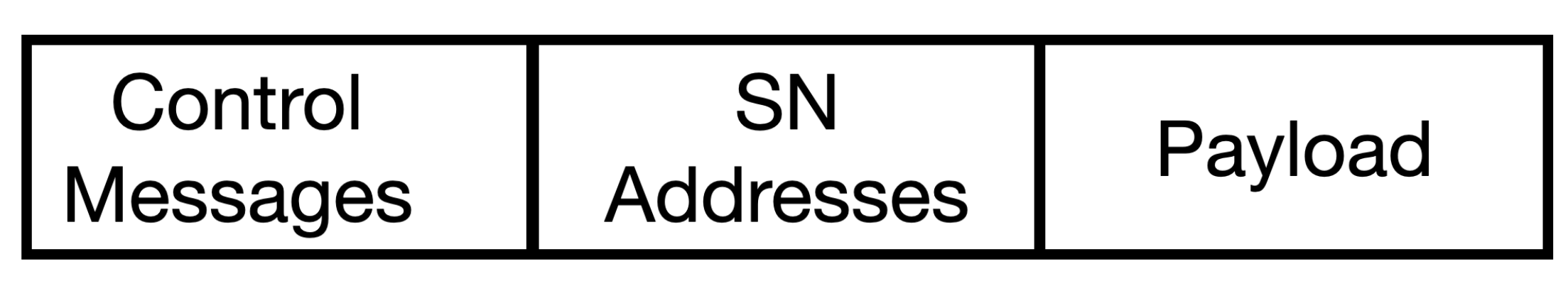}
    \caption{Physical layer payload format for the connection-based upper layer protocol.}
    \label{fig: protocol}
\end{figure}

\subsubsection{Connection-based Upper Layer Communication Protocol Design} 
The major purpose of this communication protocol is to establish reliable communication between nodes for multiple messages transmission without interruption and for the scalability of the network. Unlike the convention TCP/IP protocol stack, due to the simple network topology and the single purpose of the proposed network, we combine the networking layer and session layer functionalities into one single layer. The protocol format is shown in Fig~\ref{fig: protocol}. The first byte is the control message including the start/close a connection, acknowledgement (ACK)/negative acknowledgement (NACK) of the connection, multiple-message indicator, and so on. The second byte is the address of the SN. The payload is the transmitted data from the measurements. The detailed protocol description is as follows. In order to send data to GW, each SNs needs to first establish a connection before data is sent. Once the connection is made, the GW exclusively processes the data from the connection and denies connections attempted from other nodes. After sending data, the node will need to disconnect. If a node does not perform any activity after certain period then the GW will close the connection with the node. After sending each message, a response will be sent from the GW to the SN to indicate whether the message is received or not. If this message is not received, the SNs resend the messages after a random time during a specified interval, which doubles after each retry, up to one hour. 
This allows avoiding collisions between multiple nodes attempting to transmit simultaneously and allows for more reliable transmissions.



\subsubsection{Low Power}
Power consumption is a key part in the project since the project is implemented by low power wireless sensor networks. In order to last the longer lifetime of the system with battery power supply, an ultra-low power MCU chip and wireless transmitter is selected. Moreover, a custom PCB is designed for reducing the power consumption using the STM32L051K8 MCU as shown in Fig.~\ref{fig: MCU_pinout}. This custom PCB design allows direct inputs for power to avoid any regulator losses and avoids using unnecessary components that assume power such as LEDs. The LoRa unit takes 3.3 $V$ input, and one pin is chosen for the interrupt. The MCU provides power supply and pin connections to the LoRa module.
Sleep mode is also enabled on MCU and LoRa module when the system is not transmitting data, the sleep mode consumes much less power compared to the idle mode and the active mode.

\subsubsection{Open Source}
To ensure the implemented system is widely used by others, SPARC-LoRa is open source for future research and applications. All the system designs and codes are available at 
\cite{lora2022sensornetworks}. 


\section{Experimental results}


In this section, we will present the experimental results of SPARC-LoRa. In particular, we will test its power consumption, communication distance without or with the relay module, and the corresponding reliability. 

\subsection{Power Consumption}\label{power consumption}

As mentioned in previous sections, only the SNs requires low power consumption. Hence, we only measure the power consumption for the SNs, each of which has three stages: deep sleep mode (sleep mode), active mode, and idle mode. 
The LoRa unit's default status is sleep mode when it is powered up and will be woken up by the interrupt from processing unit to initialize the data transmission from SN to GW. The status between MCU woken up and LoRa module data transmission is named idle mode. The LoRa unit will go back to sleep mode when the transmission is completed.
Based on our measurements, the sleep mode current is approximately $180$ $\mu$A with $3.3$ V voltage supply, then, the sleep mode power consumption is $594$ $\mu$W or $0.594$ mW. The idle mode 
is $1$ second between sleep mode and active mode. 
The idle mode consumes $9.0$ mA from the measurement, with the voltage supply of $3.3$ V, the power consumption during idle mode is $29.7$ mW. When the LoRa unit is under active mode, the LoRa unit is actively transmitting data from SNs to GW, the transmission time is $12$ seconds. 
The current consumption is $75$ mA during the data transmission, the voltage supply is staying $3.3$ V. Therefore, the power consumption is calculated as $247.5$ mW on active mode. The power consumption results of SNs is summarized in Table~\ref{table: LoRa unit power consumption}.


    
    
    

\begin{table}[h]
    \centering
    \caption{LoRa Unit Power Consumption}
    
    \begin{tabular}{|c|c|}
    \hline
    Power Consumption   &    SNs \\
    \hline
    Sleep Mode          &  594 $\mu$W \\
    \hline
    idle Mode         &  9.0 mW \\
    \hline
    Active Mode         &  247.5 mW \\
    \hline
    \end{tabular}
    
    \label{table: LoRa unit power consumption}
    
\end{table}

 \begin{figure}
  \centering

  \subfigure[Grassland Testing]{\includegraphics[width=0.48\linewidth]{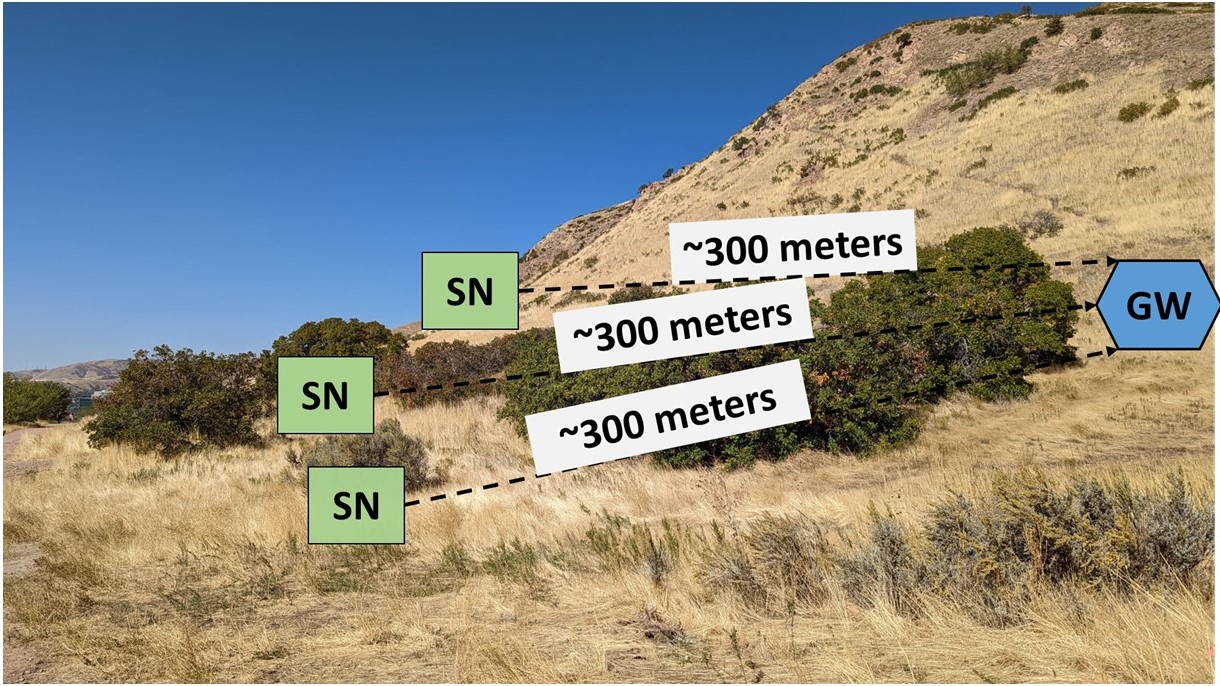}\label{fig: grassland no relay}}
  \hspace{0.2mm}
  \subfigure[Nebraska Field Testing]{\includegraphics[width=0.48\linewidth]{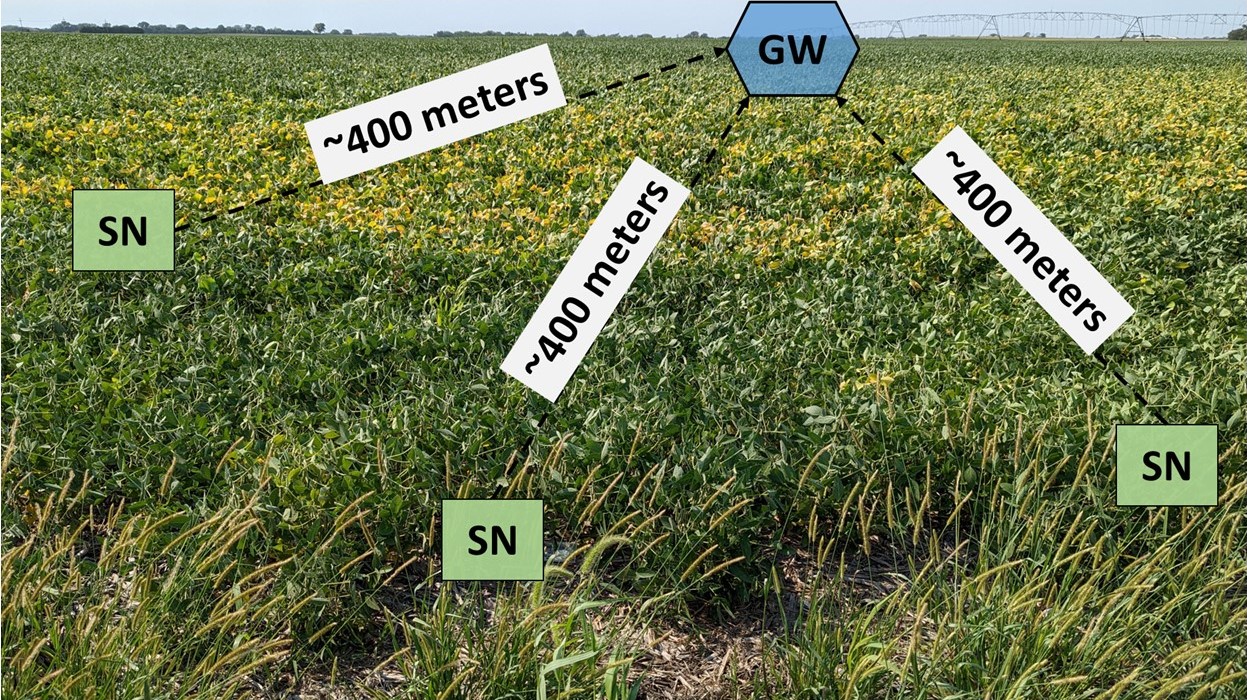}}

  \caption{Communication Testing without Relay Module.}
  \vspace{-0.3cm}
  \label{fig:Communication testing without relay module}
\end{figure}

\begin{table}[h]
    \centering
    \captionsetup{justification = centering, margin = 1 cm}
    \caption{FLR results on grassland when the GW is on the ground}
    
    \begin{tabular}{|c|c|c|}
    \hline
    Distance        & {SN (Ground)}       & {SN (1.5 meters)} \\
    \hline
    200 meters      & 0.5\%                         & 0.5\% \\
    \hline
    300 meters      & 0\%                           & 1\% \\
    \hline
    500 meters      & 0.5\%                         & 0\% \\
    \hline
    800 meters      & No data                       & No data \\
    \hline
    \end{tabular}

    \label{table: Grounded gateway data frame loss rate on grassland experiments}
    
\end{table}

\begin{table}[h]
    \centering
    \captionsetup{justification = centering, margin = 1 cm}
    \caption{FLR results on grassland when the GW is at $1.5$ meters}
    
    \begin{tabular}{|c|c|c|}
    \hline
    Distance        & {SN (Ground)}       & {SN (1.5 meters)} \\
    \hline
    200 meters      & 0\%                           &  0\% \\
    \hline
    300 meters      & 1\%                           & 0\% \\
    \hline
    500 meters      & 0\%                           & 0\% \\
    \hline
    800 meters      & 5\%                           & 3\% \\
    \hline
    \end{tabular}

    \label{table: 1.5 meters height gateway data frame loss rate on grassland experiments}
    
\end{table}

\begin{table}[h]
    \centering
    \captionsetup{justification = centering, margin = 1 cm}
    \caption{FLR results on soybean field when the GW is at $2$ meters}
    
    \begin{tabular}{|c|c|c|}
    \hline
    Distance    & {SN (Ground)}       & {SN  ($1$ meters)} \\
    \hline
    400 meters  & 11\%                          &  1.5\% \\
    \hline
    \end{tabular}

    \label{table: 2 meters heigh gateway data frame loss rate on field experiments}
    
\end{table}

\subsection{Communication without Relay Module}\label{communication without relay module}

In this section, our goal is to verify the performance of the SPARC-LoRa system under different environments including a parking lot in the University of Utah, a grassland region in Salt Lake City, Utah, and a soybean research field in the University of Nebraska-Lincoln. The metrics that we use is the transmission distance, the frame loss ratio (FLR), and the heights of SNs and GW, where the FLR is defined as the ratio between the lost data (e.g., lost data between SNs and GW, lost acknowledgement from GW to SNs, or unable to publish to online database) and the total number of transmitted data in a certain transmission period. It can be seen that the data FLR is a metric of reliability. For the data transmission in the following experiments, three SNs transmit $200$ packets continuously in each experiment. In addition, the distance between two SNs is roughly $10$ meters.



First, we did the experiment in a large parking lot in the University of Utah. In this experiment, both SNs and GW were placed on $1.5$ meters height. SNs and GW were $200$ meters apart. We tested two scenarios including line-of-sight (LOS) and none-line-of-sight (NLOS) communications.\footnote{For NLOS, there is a building between SNs and GW.} For the former, the FLR is $0\%$ or there is no data loss. In the former, the FLR is $0.5\%$. 

Second, we did the experiment in a mountainous grassland region in Salt Lake City, in particular, the Living Room trail was selected for its easy accessibility. In all the tests, the LOS is generally guaranteed between SNs and GW. However, when we put either SNs or GW or both on the ground, there are dense grass between them. The results are shown in Table~\ref{table: Grounded gateway data frame loss rate on grassland experiments} and Table~\ref{table: 1.5 meters height gateway data frame loss rate on grassland experiments}. In Table~\ref{table: Grounded gateway data frame loss rate on grassland experiments}, the GW is on the ground, while in Table~\ref{table: 1.5 meters height gateway data frame loss rate on grassland experiments}, the GW is at 1.5 meters height. It can be seen that when the GW is on the ground, the communication can be established when the distance between SNs and GW is less than or equal to $500$ meters. The communication cannot be established when the distance between SNs and GW is less than $800$ meters.\footnote{Note that this is in contrast to the LoRa's transmission distance, which can be more than $10$ miles in rural areas. However, in practice, many factors need to be considered including the heights of transmitters and receivers.} If the GW is at $1.5$ meters height, from Table~\ref{table: 1.5 meters height gateway data frame loss rate on grassland experiments}, it can be seen that the communication can be established at $800$ meters and the FLR is very low when the communication distance between SNs and GW is less than $800$ meters.

Third, we did the experiment on a soybean research field in the University of Nebraska-Lincoln. In this field, the soybean crops were approximately $1$ meter height. The density of soybean field is much thicker than the grassland. The GW was placed at $2$ meters height and the SNs were placed at $1$ meter. The result is shown in Table~\ref{table: 2 meters heigh gateway data frame loss rate on field experiments}. It can be seen that in the real soybean field, the communication can be established reliably when the communication distance is less than $400$ meters. 


\subsection{Communication with Relay Module}\label{communication with relay module}

\begin{figure}[htbp]
    \centering
    \includegraphics[width=75mm]{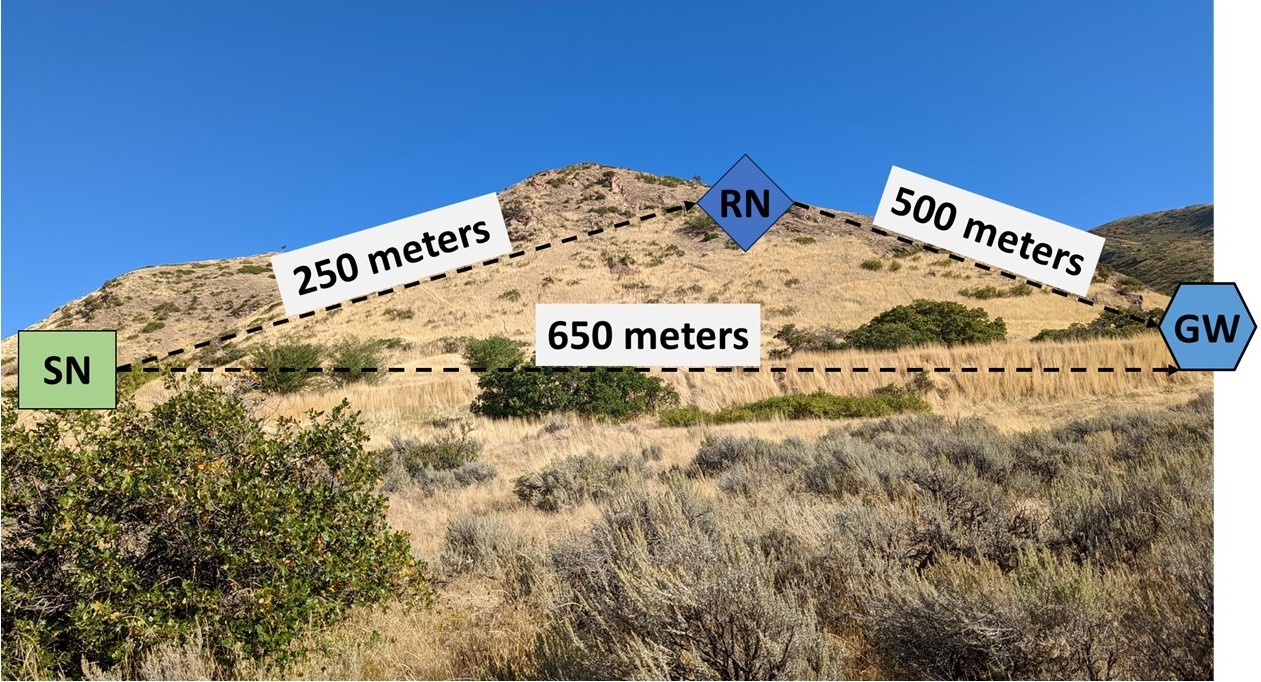}
    \caption{Relay Node Testing on Grassland.}
    \label{fig:Relay node testing in different conditions}
\end{figure}
\begin{table}[h]
    \centering
    \captionsetup{justification = centering, margin = 1 cm}
    \caption{FLR on grassland with relay module enabled}
    \begin{tabular}{|c|c|c|}
    \hline
    Distance      &   {SN (Ground)}   &      {SN ($1$ meters)} \\
    \hline
    650 meters    &          2.5\%              &            0\% \\
    \hline
    \end{tabular}
    \label{table: 1.5 meters height gateway data frame loss rate on grassland experiments with relay module}
\end{table}

    

    
    

In this section, we will present our experimental results of SPARC-LoRa when the relay module is enabled.
We did the experiment in mountainous grassland region in Salt Lake City where is the same region for the scenario where the relay module is not enable as shown in Fig.~\ref{fig: grassland no relay}. 
In this figure, SNs and GW were 
spaced out by terrain convex region and the distance between SNs and GW is $650$ meters and they are not wthin LOS. The height of GW is $1.5$ meters. In this case, the communication between the SNs and the GW cannot be established without the RN when the SNs are either on the ground or with a height of $1.5$ meters. 
When the relay module is enabled, the RN is placed on $1$ meter height. In this scenario, the distance between the SNs and the RN is $250$ meters with LOS and the distance between the RN and the GW is $500$ meters with LOS. The result is shown in Table~\ref{table: 1.5 meters height gateway data frame loss rate on grassland experiments with relay module}. 

\section{Conclusion}

In this paper, we introduced SPARC-LoRa, which is a holistic LoRa-based IoT system for agriculture applications. SPARC-LoRa satisfies all the desired requirements for such applications including scalability, low power, affordability, and reliability. In addition, SPARC-LoRa provides cloud services in order to guarantee the measured data can be viewed from any platforms or web browsers. We tested SPARC-LoRa in real fields in both Salt Lake City and the University of Nebraska-Lincoln to validate its outstanding performance. In the future, we will add other cloud services to SPARC-LoRa such as data analytics and extend it for other applications.

\section*{Acknowledgment}
This research work was 
supported by the
cooperative agreement of DE-AR0001064 of the ARPA-E
OPEN 2018 program (PM: Dr. David
Babson). Microfabrication was performed at the 
Utah Nanofabrication Facility in the University of Utah.



\bibliographystyle{IEEEtran}
\bibliography{references_d2d}

\end{document}